\documentclass[10pt]{article}
\usepackage[letterpaper]{geometry}
\usepackage{hicss}
\usepackage{times}
\usepackage[none]{hyphenat}
\usepackage{url}
\usepackage{latexsym}
\usepackage{graphicx}

\title{Dialog-based Automation of Decision Making in Processes}

\author{Bedilia Estrada-Torres \\
  Universidad de Sevilla, Seville, Spain \\
  {\underline{ iestrada@us.es}} \\\And
  Adela del-Río-Ortega \\
  SCORE Lab, I3US Institute. \\Universidad de Sevilla, Spain \\
  {\underline{ adeladelrio@us.es} }\\\And 
  Manuel Resinas \\
  SCORE Lab, I3US Institute. \\Universidad de Sevilla, Spain \\
  {\underline{resinas@us.es}} \\}

\date{}

\begin{document}
\maketitle
\begin{abstract}

The use of chatbots has spread, generating great interest in the industry for the possibility of automating tasks within the execution of their processes. 
The implementation of chatbots, however simple, is a complex endeavor that involves many low-level details, which makes it a time-consuming and error-prone task. 
In this paper we aim at facilitating the development of decision-support chatbots that guide users or help knowledge workers to make decisions based on interactions between different process participants, aiming at decreasing the workload of human workers, for example, in healthcare to identify the first symptoms of a disease. 
Our work concerns a methodology to systematically build decision-support chatbots, semi-automatically, from existing DMN models. Chatbots are designed to leverage natural language understanding platforms, such as Dialogflow or LUIS. 
We implemented Dialogflow chatbot prototypes based on our methodology and performed a pilot test that revealed insights into the usability and appeal of the chatbots developed.

\end{abstract}

\section{Introduction}
\label{sec:introduction}

In recent years, the use of artificial intelligence techniques, automated tools for the execution of tasks within business processes and the change in communication strategies between companies with customers have encouraged a large number of organizations to implement virtual assistants that provide information, solve doubts or help in the achievement of a specific task \cite{Valtolina_2108_DomainOfUse}. Often these virtual assistants take the form of chatting bots, also known as \textit{chatbots}, which are tools designed to interact with users through friendly conversations using natural language in a way that simulates interaction with a human \cite{Hussain_2019_SurveyChatbots,Valtolina_2108_DomainOfUse}. 

Chatbots have proven to be useful in a variety of areas such as healthcare \cite{Mujeeb_2017_Aaquabot}, marketing, education, business, and e-commerce \cite{Shawar_2007_Chatbotuseful}. Furthermore, they have widespread use as the digital assistants that are now on every mobile phone or on home controllers. Some authors also consider that this type of tool can go beyond its current capabilities and act as teammates of human workers in a collaborative way in complex processes \cite{Seeber_2020_chatbotsMates}.

One of the aspects that have propelled the use of chatbots is the improvement of the development of chatbots in the last years. Chatbot development platforms have abstracted away many details related to the natural language processing of a chatbot through the automated recognition of user intents, and have provided a general framework in which the conversation flow can be defined. However, a chatbot developer is still needed to implement a specific conversation flow, to deal with many low-level details about parameters or entities, to provide a set of training phrases for each of the conversation steps, and to provide a generic set of fallback options that guides users who do not know the capabilities the chatbot gives, amongst others. Implementing these tasks is usually time-consuming and error-prone even for simple chatbots \cite{Gwendal_2019_Framework}.

The hypothesis of our research is that chatbots that serve a similar purpose have many aspects on their structure and conversation flow that can be reused. Therefore, it could be possible to use these commonalities to design methodologies that partially automate the development of some particular types of chatbots. Specifically, in this paper, we exploit this idea by proposing a methodology to partially automate the development of a particular type of chatbots, namely \emph{decision-support chatbots}.

A \emph{decision-support chatbot} is a task-oriented chatbot whose purpose is to help or guide users to make decisions.
The usefulness of this kind of chatbots can be illustrated with two use cases. Let us imagine a knowledge worker that is performing a knowledge-intensive process in the context of a bank. As it is common these days, to perform her tasks, she interacts with other process participants not through email, but through a workstream collaboration tool like \textit{Microsoft Teams}\footnote{\url{https://www.microsoft.com/en/microsoft-365/microsoft-teams/}} or  \textit{Slack}\footnote{\url{https://slack.com/}} \cite{Stoeckli_2018_slackChatbots}. To perform her tasks, she needs to know the risk category of a person, which is a decision that is partially automated based on a decision table. Without leaving her collaboration environment, she asks the chatbot for the risk category of an existing customer. After a couple of interactions, in which the chatbot collects the required information such as the risk score of the customer, she obtains the answer.
The second example is related to end-users of a business process, like customers, citizens, and patients. In many cases the interaction or part of the interaction of an end-user with a business process involves asking her for information to get a decision. For instance, during the COVID-19 outbreak, many healthcare services around the globe have implemented apps that ask people about their symptoms and their contacts with infected people, such as \cite{Brandtzaeg_2017_WhyUseChatbot}, providing information to end users and helping to decrease the medical workload. Based on their answers, the app may advice about the best next step. These interactions can be easily performed by a decision-support chatbot, as shown in \cite{web_chatbot_vasco}, removing the need to create dedicated apps for that, and making them more accessible to all kinds of users.

Our methodology removes or systematizes many of the low-level tasks that chatbot developers need to perform to implement a decision-based chatbot like the ones described above, letting the developer focus only on the definition of the decision using decision tables. In order to evaluate our approach, chatbot prototypes were implemented using Dialogflow\footnote{\url{https://dialogflow.cloud.google.com/}}, a chatbot development framework. A set of pilot users from academia and industry interacted with these prototypes so that we gathered insight into the usability and appeal of the bots that were developed in this way.

The rest of this paper is structured as follows. 
Section \ref{sec:motivating_example} introduces a motivating example for this work.
Sections \ref{sec:related_work} and \ref{sec:key_concepts} describe related work and key concepts of chatbots, respectively. 
The proposed methodology is described in Section \ref{sec:metodology}. 
The implementation of chatbots, the evaluation of the proposal and its limitations are presented in Section \ref{sec:evaluation}. 
Finally, Section \ref{sec:conclusions_future_work} concludes the paper and describes challenges for future work.
\section{Motivating Example}
\label{sec:motivating_example}

To illustrate the methodology we propose, we use the example based on a bank scenario, in which it is necessary to decide the \textit{risk category} of a person (customer). The risk category can be \textit{high, medium} or \textit{low}, depending on three criteria: if the person is an \textit{existing customer} of the bank, what the application \textit{risk score} that she provides is and what the assigned \textit{credit score} is. 

The automation of decisions requires the organization to have previously identified and defined decisions. In this paper, we use the Decision Model and Notation (DMN) \cite{OMG_DMN_2019} to illustrate and model the decisions. We have chosen it because it is a well-known standard that provides a notation for describing and modeling repeatable decisions like this one in a readily understandable way for analysts, technical developers and business users. Specifically, our contribution focuses on decision tables, which are used for the definition of expressions, calculations, if/then/else logic, amongst other. However, we must emphasize that our approach is independent of the specific notation used to represent decision tables.

\begin{figure}[thb]
	\centering
	\includegraphics[scale=0.35]{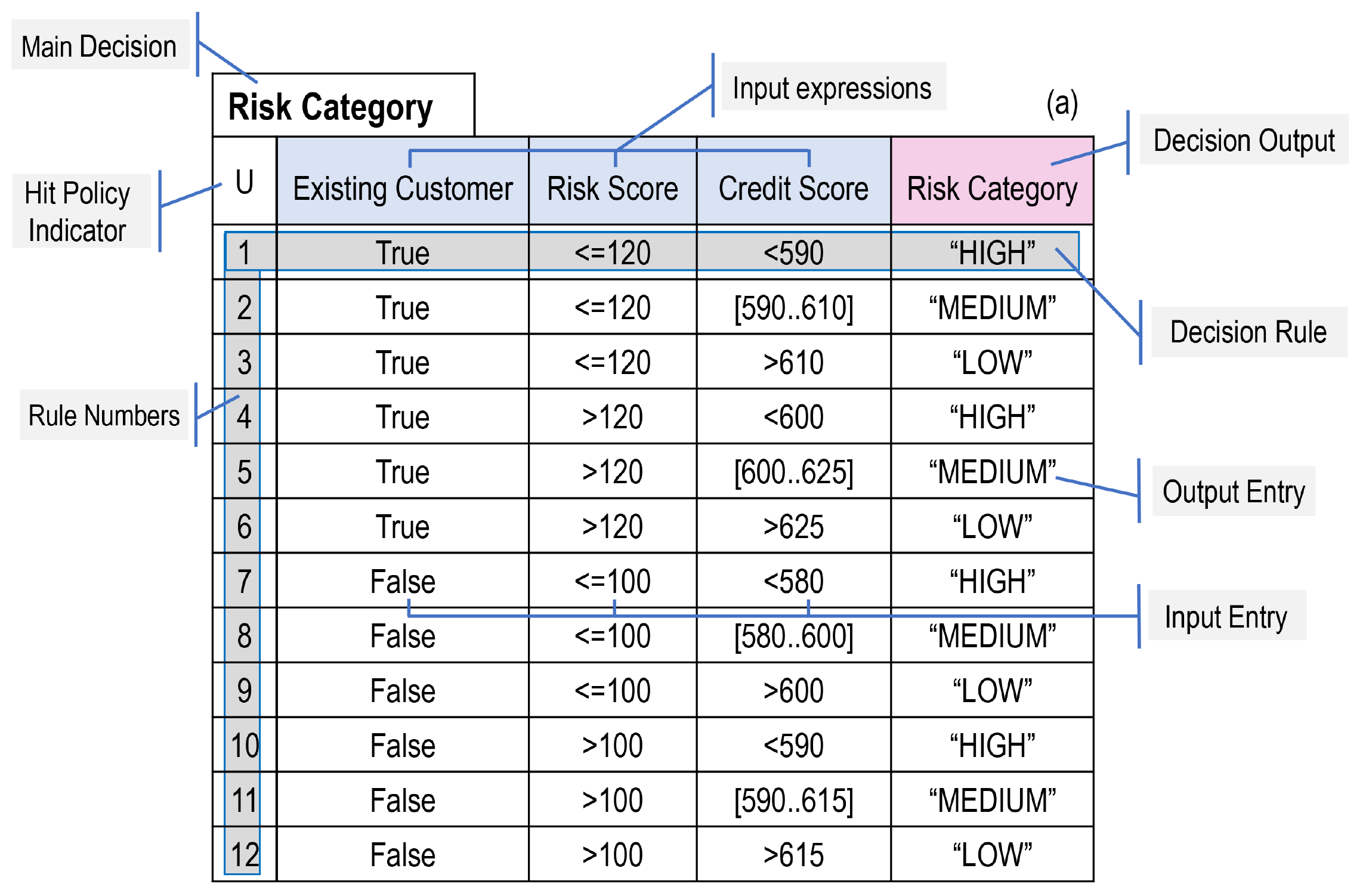}
	\caption{Example of a DMN Decision Table (Taken from \cite{OMG_DMN_2019})}
	\label{fig:dmn_table_1}       
\end{figure}

\begin{figure}[thb]
    \centering
	\includegraphics[trim={0cm 0cm 0cm 0cm}, clip,width=8cm]{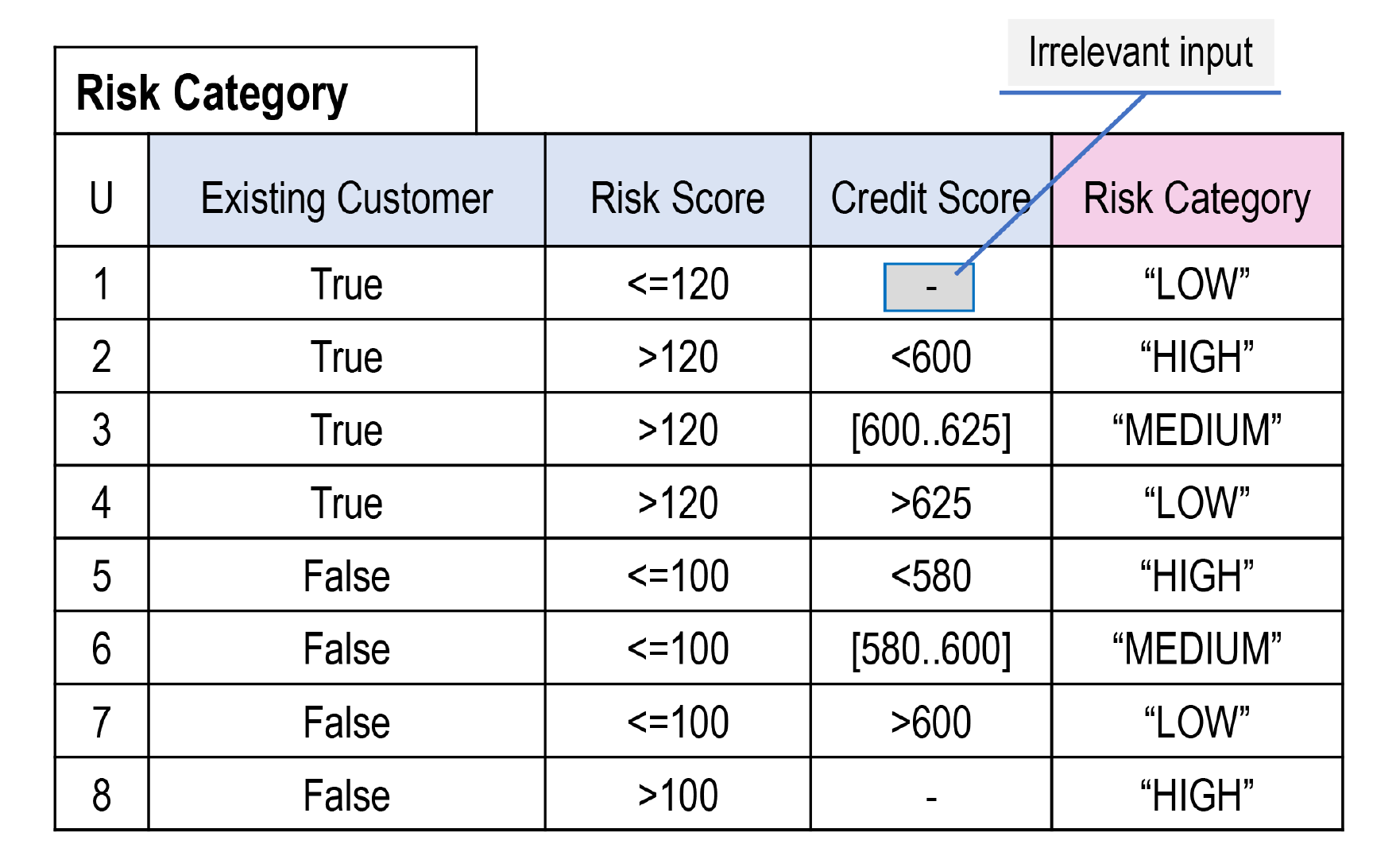}
	\caption{DMN Decision Table with irrelevant inputs}
	\label{fig:dmn_table_2}       
\end{figure}

A decision table, as shown in Figures \ref{fig:dmn_table_1} and \ref{fig:dmn_table_2}, define the possible combinations between the three criteria required to determine the risk category. 
More specifically, the \textit{decision} (Risk Category) is resolved using a set of \textit{decision rules} (rows from 1 to 12 for Figure \ref{fig:dmn_table_1}, and from 1 to 8 for Figure \ref{fig:dmn_table_2}). Each \textit{decision rule} is composed of a set of \textit{input} (3) and \textit{output entries} (1) and is identified by a \textit{rule indicator}. The \textit{hit policy indicator} (U, unique) defines the output values to be selected in the decision-making process.

Decision tables can be defined horizontally or vertically. In these examples, they are all vertical, where each column with \textit{input expression} (Existing Customer, Risk Score and Credit Score) represents a type of input entry; and the last column (Risk Category) represents the \textit{decision output}. 
All decision table entries in Figure \ref{fig:dmn_table_1} are required. The entries in Figure \ref{fig:dmn_table_2} are slightly different because some of them are not significant (wildcards, ``-''), the \textit{credit score} in rules 1 and 8. 
Due to the fact that there is a large number of proposals aimed at checking the quality of definition of the decision tables in DMN, such as the overlap between rules \cite{Batoulis_2018_DMNDecision}, or to verify the consistency, correctness and coverage of decision rules \cite{Huysmans_2007_consistency,Qian_2012_OrderedDT,Calvanese_2018_DMNCorrectness}, in this article, we assume the decision tables used to build a chatbot were built correctly. 

\begin{figure}[thb]
    \centering
	\includegraphics[trim={0cm 0cm 0cm 0cm}, clip,width=8cm]{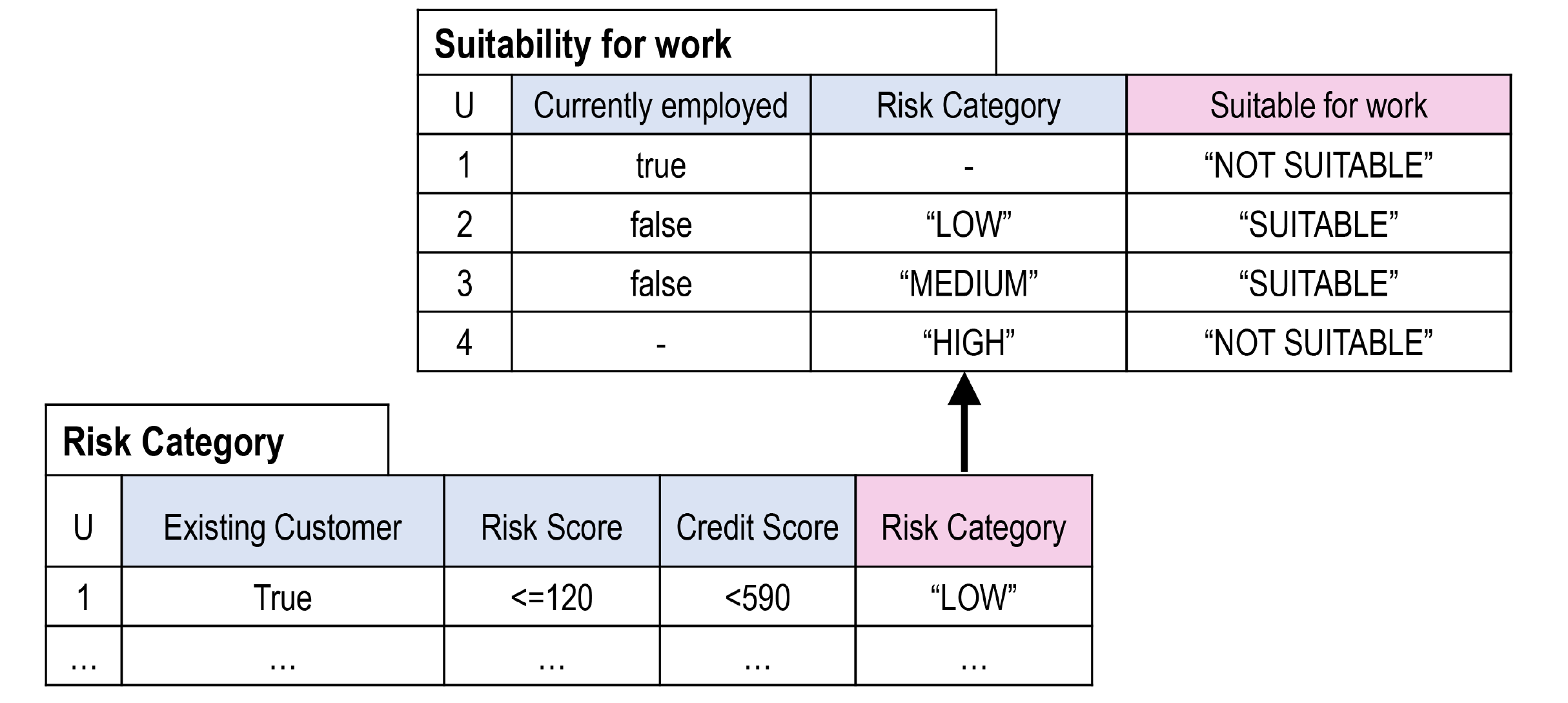}
	\caption{DMN Decision Table with a decision as input value}
	\label{fig:dmn_table_3}       
\end{figure}

\begin{figure}[thb]
    \centering
	\includegraphics[trim={0cm 0cm 0cm 0cm},clip,width=8cm]{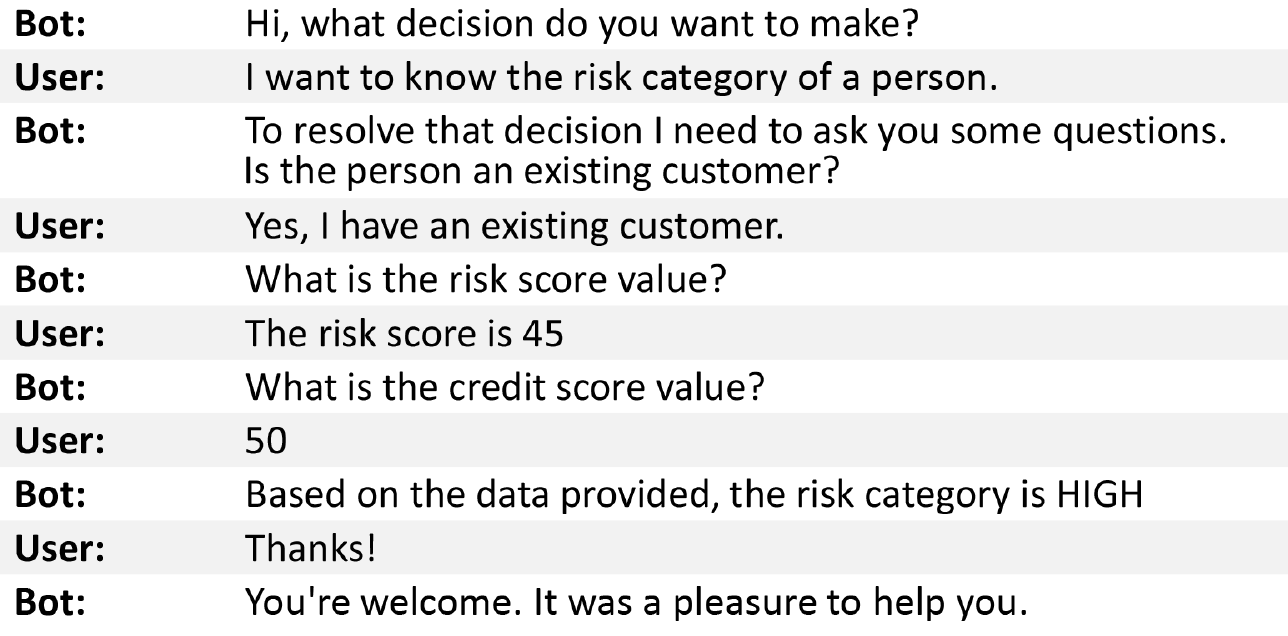}
	\caption{Interaction between a Decision-support chatbot and a user}
	\label{fig:conversation_bot}       
\end{figure}

The aim of this research is to generate a decision-support chatbot that interacts with a user in a specific domain defined by a decision table. Figure \ref{fig:conversation_bot} shows an example of the interaction between a chatbot and a user to perform the decision. The chatbot asks for the input entries of the decision table, except when the input entry is defined as ``-''. In that case the chatbot must determine whether it is necessary or not to ask for the attribute value. For example, if after an interaction the user indicated the attributes \textit{Existing customer = True} and \textit{Application risk score = 50}, it is not necessary to ask for the \textit{Credit score} value, because independently of the value provided by the user, the output value will be ``LOW'' (see Figure \ref{fig:dmn_table_2}).

In addition, our proposal also allows and guides the development of chatbots from decision hierarchies. That is, decisions where  value of one or more of the inputs are the result of a previous decision. 
For example, we could define another decision table (see Figure \ref{fig:dmn_table_3}) to decide whether a person is suitable for a job in an organization based on the risk category obtained from the other table.

\section{Related Work}
\label{sec:related_work}

Chatbots interact with the user simulating and reproducing conversations turn by turn using natural language through the exchange of written messages or voice commands \cite{Valtolina_2108_DomainOfUse}. 
Based on the classification of Hussain et al. \cite{Hussain_2019_SurveyChatbots}, here we focus on task-oriented and text-based chatbots. 

Chatbots are considered useful for the improvement of productivity and the speed and ease of use of their interfaces \cite{Brandtzaeg_2017_WhyUseChatbot}. 
They have been widely used in contexts as varied as healthcare \cite{Mujeeb_2017_Aaquabot}, e-commerce \cite{Cui_2017_ChatbotECommerce}, customer service \cite{Xu_2017_CbotCustomerService}, marketing \cite{Kaczorowska_2019_ChatbotMarketing} among others \cite{Shawar_2007_Chatbotuseful}, or modern assistants like \textit{Siri}, \textit{Alexa} or \textit{Google Assistant}.

Gwendal et al. \cite{Gwendal_2019_Framework} highlighted the need for in-depth knowledge of the tools for deploying which increases costs. They proposed a framework for defining chatbots, but one that requires human intervention. Our proposal could complement this framework by providing all these elements for decision-support chatbots. 
On the other hand, Syed et al. \cite{Syed_2020_RPA} identified challenges related to RPA tools. 
Among these challenges are the lack of methodological support for their implementation, as well as the systematic design, development and evolution. With our proposal we address both, since we propose a systematic solution for the development of chatbots in a semi-automatic way, with the aim of reducing development time and effort.   

After reviewing the literature we have not found any work that addresses the creation of chatbots based on decision models like  DMN decision tables. However, there are solutions to implement task-oriented chatbots in related domains.
In \cite{Perez_2018-DecisionChatbots}, authors proposed a chatbot for the collaborative creation of models based on conversations. And in the context of business processes, L\'opez et al. \cite{Lopez_2019_ProcModelChatbot} described a methodology and a prototype to transform a BPMN model into a chatbot based on AIML language\footnote{\url{http://www.aiml.foundation/}}. Although its evaluation revealed shortcomings in the fluency and certainty of the chatbot responses, it also shows the high potential of the proposal and the multiple possibilities of extension. 
\section{Key Concepts of Chatbots}
\label{sec:key_concepts}

Chatbot design typically relies on parsing techniques, pattern matching strategies and Natural Language Understanding (NLU) to process user inputs. The latter has become the dominant technique thanks to the popularization of libraries and cloud-based services such as Dialogflow, \textit{wit.ai} or LUIS, which rely on Machine Learning techniques and Natural Language Processing techniques to understand user input \cite{Gwendal_2019_Framework}. According to \cite{Gwendal_2019_Framework}, a NLU chatbot, also known as \textit{agent} \cite{Valtolina_2108_DomainOfUse}, contains a recognition engine that matches user \textit{inputs} with \textit{intentions (intents)} during a conversation turn. It also contains an execution component capable of executing \textit{actions} for each intent, such as \textit{responses} to the user.

\textit{Intentions} or \textit{Intents} are defined through \textit{training phrases}, which are input examples that allow the recognition engine to identify the different phrases that a user can utilize to express an intention. 
For instance, an intent can be defined to recognize that the user wants to perform a decision. Some training phrases for such an intent could be  ``I want to determine the risk category,'' or ``What is the risk category of an existing customer with a risk score of 35.'' In each \textit{training phrase}, concrete values can be recognized. These values are called \textit{parameters}. For instance, in the second example, there are two parameters, namely: existing customer equals to true and risk score equals to 35. The type of the parameters is defined by a specific structure called \textit{entity}, which determine how data from a user input is extracted. NLU platforms provide predefined entities that match many common types of data like numbers, dates, times, colors, or e-mail addresses. In addition, it is possible to define custom entities for enumerated values. A custom entity is composed of a set of entries. Each entry is made up of a reference value and a set of synonyms for that reference. For example, if we define a ``boolean'' entity, two entries are required: \textit{true} and \textit{false}. For the \textit{true entry}, the values \textit{yes, ok, correct}, could be synonyms for it. 

Each intent can be associated to \textit{input} and \textit{output contexts} to control the flow of the conversation. Contexts are identified by a string like \textit{awaiting risk score}. When an intent is matched, any configured output context for that intent becomes active. While any contexts are active, the chatbot is more likely to match intents that are configured with input contexts that correspond to the currently active contexts. In addition, contexts store the parameter values provided by a user in one intent so that it can be used in other intents. 

\begin{figure}[thb]
	\centering
	\includegraphics[trim={0cm 0cm 0cm 0cm},clip,width=8cm]{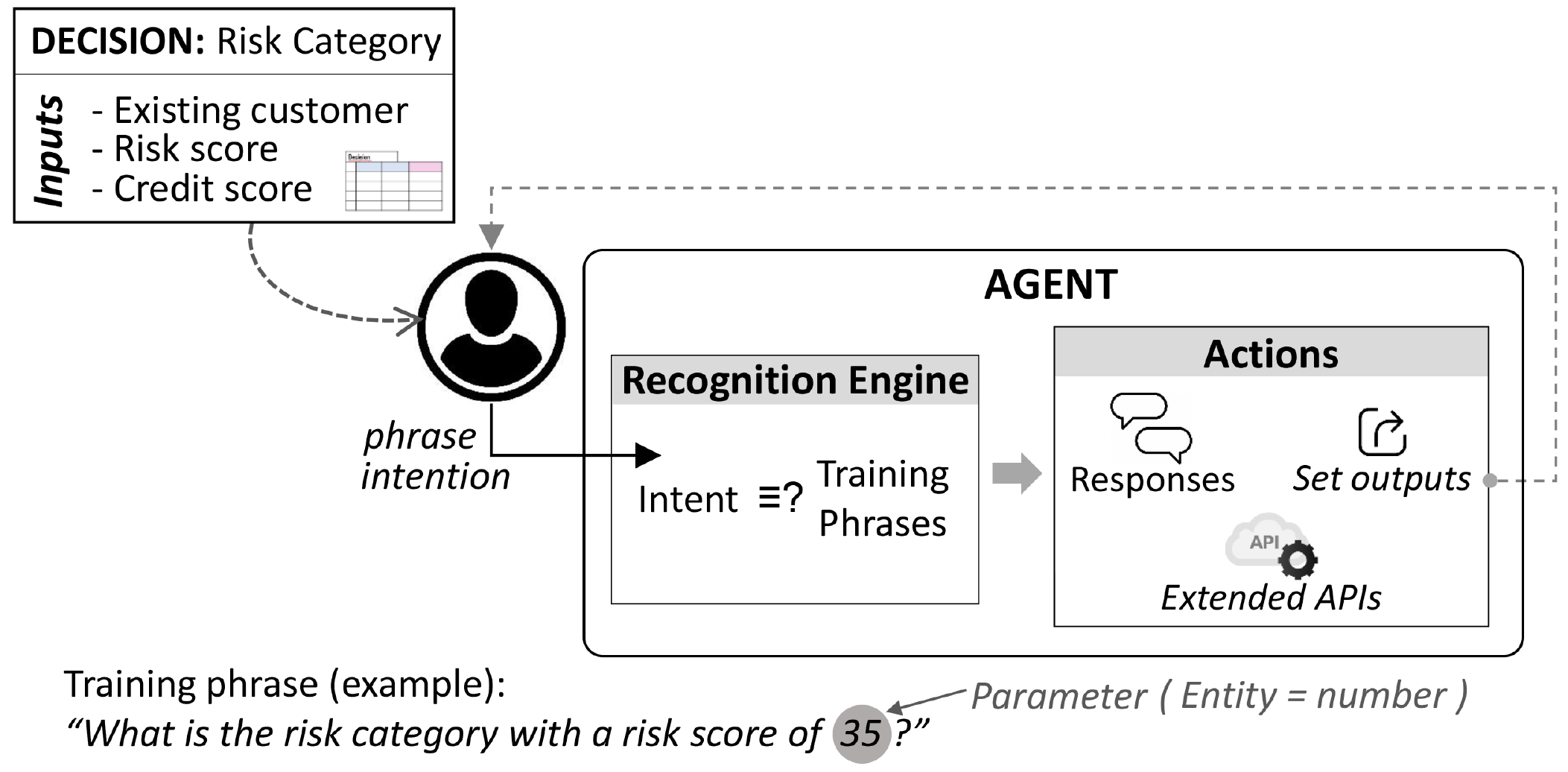}
	\caption{Interaction between key concepts of chatbots}
	\label{fig:key_concepts}       
\end{figure}

Finally, for each intent that is recognized, one or several actions can be executed. These actions include sending a response to the user, setting an additional output context, or invoking an external API, amongst others. Figure \ref{fig:key_concepts} represents interactions between the key concepts of chatbots described above.

\section{Methodology to Build Chatbots from Decision Tables}
\label{sec:metodology}

\begin{figure}[thb]
	\centering
	\includegraphics[trim={0cm 0cm 0cm 0cm},clip,width=8cm]{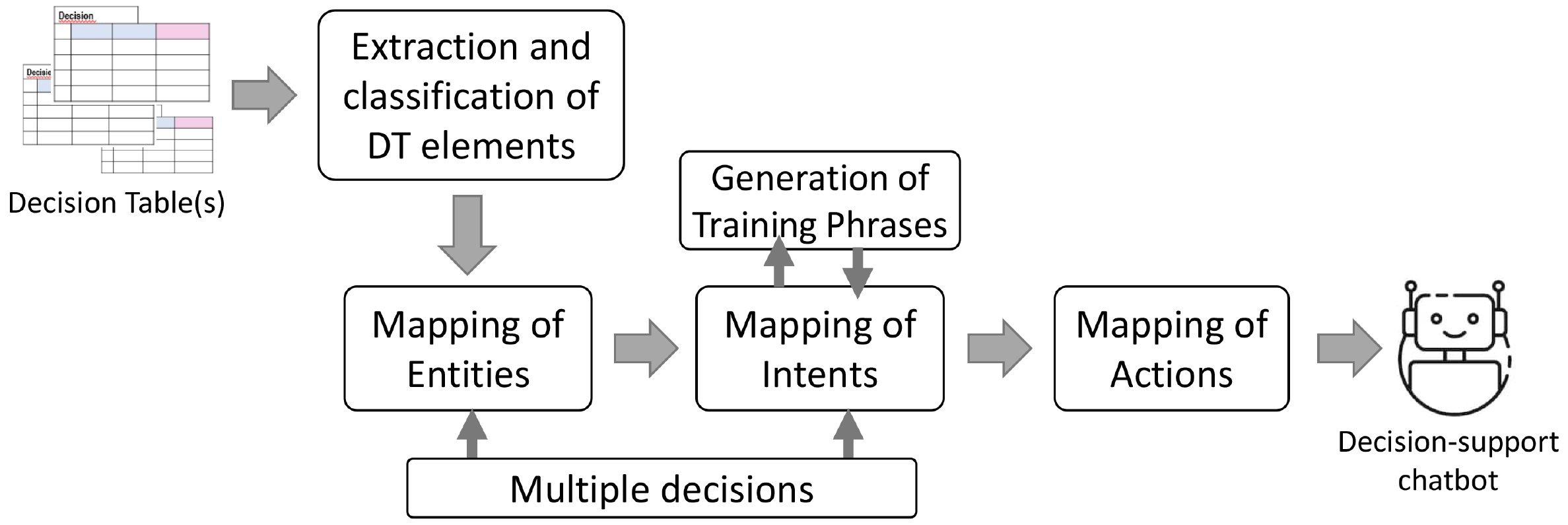}
	\caption{Methodology to Build Chatbots from Decision Tables.}
	\label{fig:metodology_diagram}       
\end{figure}

In this section, the methodology we propose to build chatbots from decision tables is described. 
The methodology takes as input a decision table or a hierarchical definition of decision tables.  
Figure \ref{fig:metodology_diagram} describes the main steps of the methodology. In the following subsections we describe how each of these steps should be performed in order to build a chatbot that meets the following three conversational requirements:

\begin{itemize}
    \item \textbf{R1:} The information required to make the decision can be provided in any order. In other words, the user does not have to learn a predefined structure that has to be used to provide the information to the chatbot.
    \item \textbf{R2:} The user can provide several input values at the same time, even in the first interaction to improve efficiency, especially for advanced users. This means that, for our example, the chatbot must be able to deal with phrases like ``What is the risk category of an existing customer with a risk score of 35.'' By doing so, the chatbot does not have to ask again about the risk score and whether it applies to an existing customer, lowering the number of interactions with the user and making the conversation more human-like.
    \item \textbf{R3:} Only information that is necessary should be asked to the user. This means that in a chatbot developed for the decision table in Figure \ref{fig:dmn_table_3}, if the user said that \textit{currently employed} is true, the \textit{risk category} should not be asked because its value is irrelevant for the decision.
\end{itemize}

\subsection{Extraction and Classification}
\label{subsec:methodology_extraction}
As a first step, we propose to identify and extract all the DMN elements of the decision model to be mapped with entities, intents, parameters, contexts, actions and training phrases. The main decision table and the hierarchy of decisions should be identified, in case there is more than one decision table, the inputs required for each decision table and the input type for each expected value. These data will be used in the following phases.

\subsection{Mapping of Entities}
\label{subsec:methodology_m_entities}

To build a chatbot several entities can be needed. In this step we propose that an entity called \texttt{ent\_<inputname>} must be created for each input whose type is not supported by the system entities of the NLU platform. In particular:
\begin{itemize}
    \item If the input type is boolean, the entity has two \textit{entries}: \emph{true} and \emph{false}. In addition, several synonyms for them must be defined. These synonyms include the typical ``ok'' and ``yes'' (and their false counterparts), but also it is considered as true the name of the input and as false its negation in several forms. For instance, for the input \emph{existing customer}, an entity \texttt{ent\_existingcustomer} is created with two entries \emph{\{True, False\}}, where \emph{synonyms(True)=\{yes, an existing customer, with an existing customer\}} and \emph{synonyms(False)=\{no, non existing customer, not an existing customer, without an existing customer\}}
    
    \item If the input type is an enumerated data type, the entity has one entry for each of the enumerated values of the data type. Here, no synonyms are provided by default. However, the chatbot developer can provide them if it makes sense for the domain. 
\end{itemize}

\subsection{Mapping of Intents}
\label{subsec:methodology_m_intents}

According to their function, and with the intention of fulfilling the conversational requirements, we propose to classify intents into three types: \textit{decision intents}, \textit{input intents} and \textit{support intents}. Each one is described below. 

\subsubsection{Decision Intents. }
\label{subsubsec:decisionIntents}

A decision intent (\texttt{<decisionname>\_intent}) is created for each $DT$ in the DMN model. Their main goal is to identify the decision that the user wants to perform. This means it must recognize phrases that convey this intent like ``I want to determine the risk category,'' or ``I want to know a risk category,'' or maybe directly ``risk category.'' In addition, according to requirement R2, the decision intent must also be able to gather as many information from that user input dealing with phrases like ``What is the risk category of an existing customer with a risk score of 35.'' As a consequence, one parameter (\texttt{p\_<inputname>}) for each input of the decision table is added to the intent. The type of each parameter is either a system entity related to the type of the input or the custom entity already defined for the input. Finally, decision intents have no input context and one predefined output context that represents the chosen decision. However, other output contexts are added during the action phase as described in the next section. 

\subsubsection{Input Intents.} 
\label{subsubsec:input_intents}

An input intent (\texttt{<inputname>\_intent}) is created for each $input$ in the DMN model. Their main goal is to gather information about each of the inputs after a question made by the chatbot (e.g., ``what is the credit score?''). However, like in the decision intent, it is necessary to provide a mechanism to capture information about other inputs that the user might include together with the response. Therefore, the phrases that need to be recognized are like ``it is 37'',  or ``the credit score is 37 and the risk score is 25.'' As a consequence, again one parameter (\texttt{p\_<inputname>}) for each input of the decision table is created. The only difference is that the parameter that corresponds to the intent associated with the same input will be defined as $required$. If a \textit{required parameter} is not provided, the intent cannot continue with the normal conversation flow and will request this parameter from the user. For example, in the \texttt{existing customer\_intent}, \texttt{p\_existingcustomer} is $required$ and \texttt{p\_riskscore} and \texttt{p\_creditscore} are $optional$. Input intents have one \textit{input context} \texttt{awaiting\_<inputname>}, which is activated to signal the moment in which that intent is requested within the conversation flow. 

\subsubsection{Support Intents. }
\label{subsubsec:support_intents}

Their goal is to make communication with the user more fluid and friendly and provide help if necessary. Unlike decision intents and input intents, they are independent of the DMN model and, hence, they can be reused in all chatbots without any change. In our proof of concept, we have included four intents with the following purposes:
\begin{itemize}
\item To manage when an expected value is not provided in the user's phrase. 
\item To recognize a user greetings and start the conversation by asking which decision she wants to use.
\item To recognize a goodbye or thank you from the user to end the conversation.
\item To recognize when the user is asking for help on how to use the chatbot. 
\end{itemize}

Other intents can be added if necessary. For instance, one could add an intent that recognizes when a user asks for the inputs for which he or she has already provided a value, or all of the inputs for which values must be provided. In fact, having the possibility of reusing these intents in all decision chatbots is a very appealing feature of our approach. 

\subsection{Generation of Training Phrases}
\label{subsec:methodology_m_training_phrases}

Each intent definition contains a set of training phrases, which are input examples used to detect which intent the user refers to. These training phrases also detail how to extract the information to fill the parameters of the intent from the user's message. For instance, the sentence ``I want to determine the risk category of a non existing customer with a risk score of 90'' should match with the main decision intent and has to fill two parameters, namely \emph{existing customer}, which evaluates to \emph{false}, and \emph{risk score}, which evaluates to 90. 

The generation of training phrases is one of the most relevant, but also more time-consuming tasks while developing a chatbot based on NLU. In our approach we provide a semi-automatic way to generate these phrases based on natural language generation (NLG). Traditionally, NLG has been used in chatbots to generate its responses to the users. However, with the advent of NLU chatbots, new NLG tools have been particularly designed for building training phrases for chatbots. The reason is that this task is slightly different from other NLG tasks because the goal is not to build a set of phrases intended for humans, but instead to build a dataset for training the chatbot. Therefore, it is not strictly necessary that the resulting phrases are fully syntactically correct because the NLU layer already deals with this aspect. Furthermore, the natural language generation must be designed to provide a wide variety of examples. There are several tools that can be used for this task. The approach followed by them is to build a generation specification that define the patterns of text that are used to create the dataset. In this paper, we use the open source project \textit{Chatito}\footnote{\url{https://rodrigopivi.github.io/Chatito/}}.

\begin{figure}[thb]
    \centering
	\includegraphics[trim={0cm 0cm 0cm 0cm},clip,width=7cm]{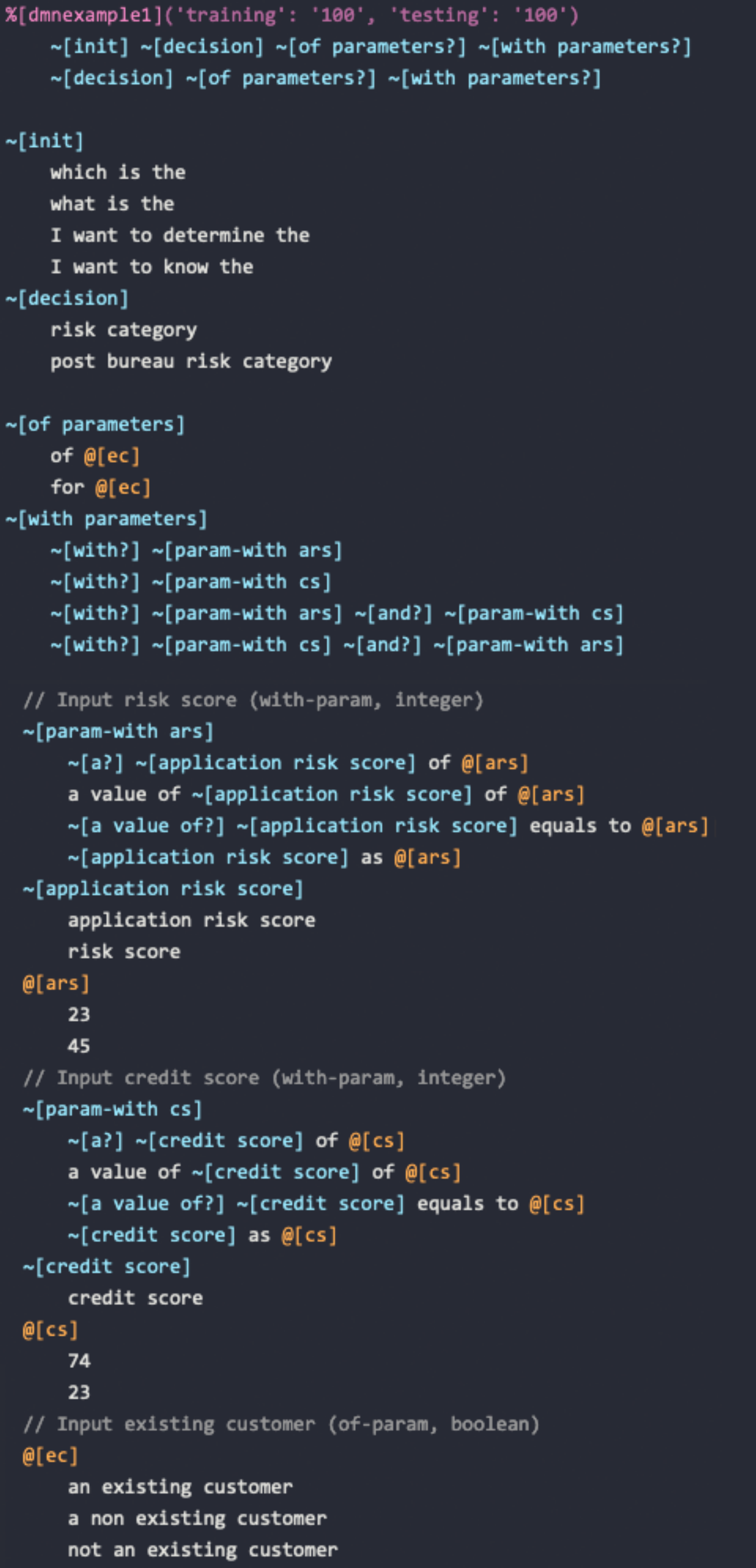}
	\caption{Natural language generation specification.}
	\label{fig:chatito}       
\end{figure}

Figure \ref{fig:chatito} depicts an extract of such a specification for the example chatbot. It includes an intent entity called \texttt{dmnexample1} that can be generated from two alternative sentences. Each of the sentences refers to alias entities like \texttt{init} or \texttt{decision} that provide alternatives for the phrase generation. Alias entities can be made optional by adding a question mark at the end of the alias name like in \texttt{\~{}[of parameters?]}. Finally, slot entities, which represent parameter values, are annotated with \texttt{@}. 
The procedure to generate training phrases requires the DMN model together with some hints about how to deal with some input parameters are processed and used to automatically build a specification for each of the intents of the chatbot. The user can then optionally refine the specification obtained automatically in the previous step. Finally, the NLG tool is used to generate the set of training phrases that will be used to train each intent in the NLU chatbot. Next we focus on how the first step is performed.

\subsubsection{Generation for Decision Intents. }
\label{subsubsec:generation_decision_intents}

The pattern for decision intents is represented by an optional, initial (\texttt{[init]} phrase like ``I want to know the,''; the name of the decision (\texttt{[decision]}), which can be either the output or the name of the decision in the DMN model, and two additional patterns to include parameters, namely: \emph{of-params}, and \emph{with-params}, although more patterns could be easily added following a similar approach. \emph{Of-params} include inputs that can be expressed better using the preposition \emph{of} like \emph{existing customer} in our example. On the contrary, \emph{with-params} include inputs that can be more naturally expressed using the preposition \emph{with} like \emph{risk score} or \emph{credit score}. The information of whether a parameter is an \emph{of-param} or a \emph{with-param} must be provided by the chatbot developer. If no information is provided, all parameters are considered to be \emph{with-params}. 
Regardless of the type of pattern, a slot entity is created for each input. The values of the slot entity depends on the type of the input and include the domain of the input or a subset if the domain is infinite like a number or a date or a string. The only exception is if the type of the input is a boolean. In that case, if the input is a \emph{with-param} it adds a second type of slot entity for the input, which includes the name of the input with ``with'' and ``without'', i.e., ``with existing customer'' and ``without existing customer.'' On the other hand, if the input is an \emph{of-param} there is only one value entity whose values are the name of the input and its negation: ``an existing customer,'' ``a non existing customer,'' and ``not an existing customer.'' 

Having the slot entities defined, the pattern of \emph{of-params} include the preposition \emph{of} and k-permutations of the slot entities created for each input with $k=\{1,\ldots,n\}$, where $n$ is the number of inputs. Furthermore, if the input is of a generic type like numbers, dates or strings, the name of the input is added after the slot entity created. If the number of permutations is too large, then only all 1-permutations, one n-permutation and a random subset of the other k-permutations are chosen. This can be done because we are only creating training examples, not defining the whole set of possible phrases the chatbot has to recognize. Regarding \emph{with-params}, the pattern follows the same approach, but starting with the preposition \emph{with}. However, unlike in the previous case, the slot entity is not used directly, but an alias entity that provides different alternative ways in which the value of the attribute can be provided. Some examples are: ``a credit score of,'' ``a value of credit score of,'' or ``credit score as.''

The permutations used to generate the training phrases for the parameters allows the user to provide the information about the different inputs in any order following requirement R1. 
As we said before, once the phrases are generated, the user can refine them if desired.

\subsubsection{Generation for Input Intents. }
\label{subsubsec:generation_input_intents}

The pattern for input intents is composed by an alias entity that represents the answer to the question (\texttt{[answer]}) and, optionally, some additional parameters. The former includes options like providing directly the slot entity, or surrounding it with some additional text like ``it is'' or ``the \emph{$\langle$parameter name$\rangle$} is.'' Because of the way the questions are made, in this case there is no difference between \emph{of-params} and \emph{with-params}.

Concerning the additional parameters, they are added following the same approach as before, although with some slight changes. Since the context of this phrase is slightly different than the decision intent, a new alias entity is added that allows the definition of \emph{of-params} using the pattern ``it is'', and \emph{with-params} using the pattern ``the \emph{$\langle$parameter name$\rangle$}.'' For instance, after the question ``what is the credit score?'', a possible answer could be: ``it is 37 and the risk score is 47.'' Furthermore, the same entities for specifying \emph{of-params} and \emph{with-params} than in decision intents is also included, although in \emph{of-params}, ``of'' is changed by ``it is.'' This means that another possible answer could be: ``37, and it is an existing customer with a risk score of 25.'' Obviously, the input that appears in the question is removed from all these entities. 

Finally, note that some heuristics could be added to improve the automated process for some common types of inputs. For instance, if an input is \emph{Age}, the generation could be adapted so that instead of creating phrases that include \emph{a 23 age}, it creates phrases that include \emph{23 years old}.

\subsection{Mapping of Actions}
\label{subsec:methodology_m_actions}

\begin{figure}[thb]
	\centering
	\includegraphics[trim={0cm 0cm 0cm 0cm},clip,width=8cm]{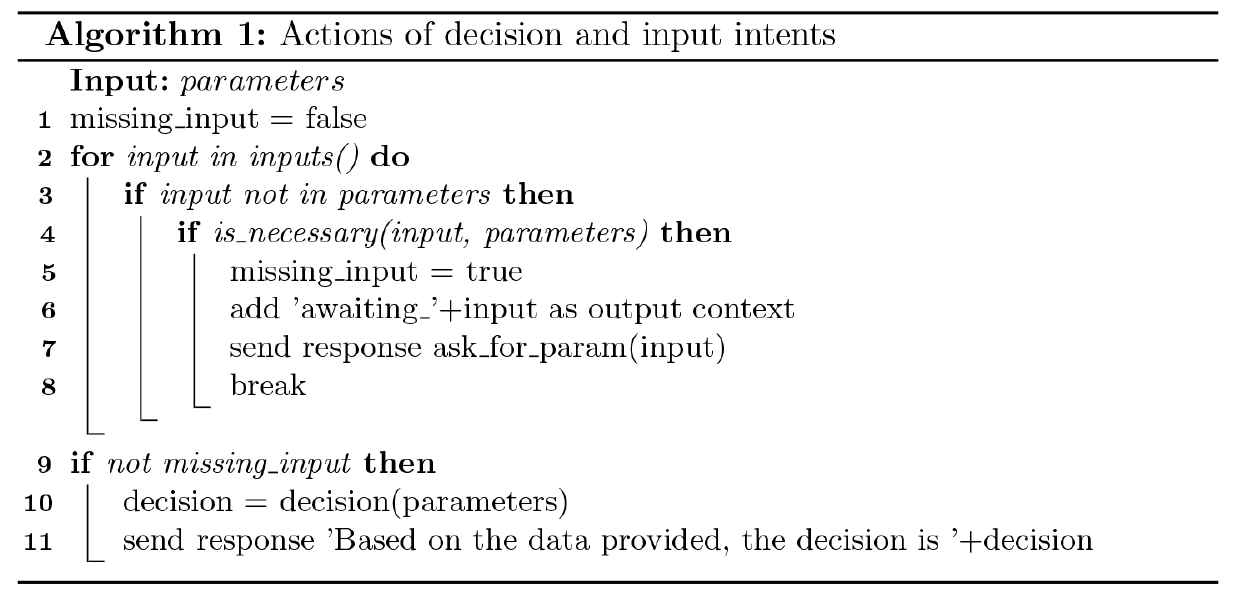}
	\caption{Actions of decision and inputs intents}
	\label{alg:required_inputs}       
\end{figure}

Each intent has an associated action that is performed when the intent is recognized. The most straightforward action is a canned response provided by the chatbot. This is the action used with the support intents. For example, a farewell intent may answer \textit{You're welcome}, \textit{Come back soon}, etc. However, the action that needs to be implemented for decision and input intents is more elaborated. 
Our proposal uses the algorithm shown in Figure \ref{alg:required_inputs}, which depicts the response action, every time a question turn occurs in which the user answers. 
It assumes we have three functions available for the decision at hand: \texttt{inputs()}, which returns all the inputs of the decision; \texttt{decision(parameters)}, which returns the decision for the given set of parameters, where \texttt{parameters} is a map that assigns a value to each input, and \texttt{is\_necessary(input, parameters)}, which returns whether the given input is necessary given the current values assigned to the parameters. For instance, in the example of Figure \ref{fig:dmn_table_3}, \texttt{is\_necessary(Risk Category, Currently employed = true)} would evaluate to false because the value of \emph{Risk Category} does not affect the decision, whereas \texttt{is\_necessary(Risk Category, Currently employed = false)} would evaluate to true. The execution of the functions is carried out in an external service.

Following Algorithm 1, it receives as input the parameters recognized in the intent and iterates over the inputs of the decision (line 2). Then, it is checked if a parameter value has already been provided for each input (line 3). If the input is missing, it is checked whether with the current parameters, the missing input is really necessary by means of function \texttt{is\_necessary} (line 4). Therefore, only required information is asked to the user as imposed by requirement R3. If this is the case, an output context is created and activated with the name of the missing input and a response like ``What is the \textit{Risk Score} value?'' is sent requesting the missing value. This is performed by function \texttt{ask\_for\_param} (lines 5-8).
If the parameters have been provided for all the required inputs, the decision is made and a response is sent including the decision result (lines 9-11).

\begin{figure}[thb]
	\centering
	\includegraphics[trim={0cm 0cm 0cm 0cm},clip,width=8cm]{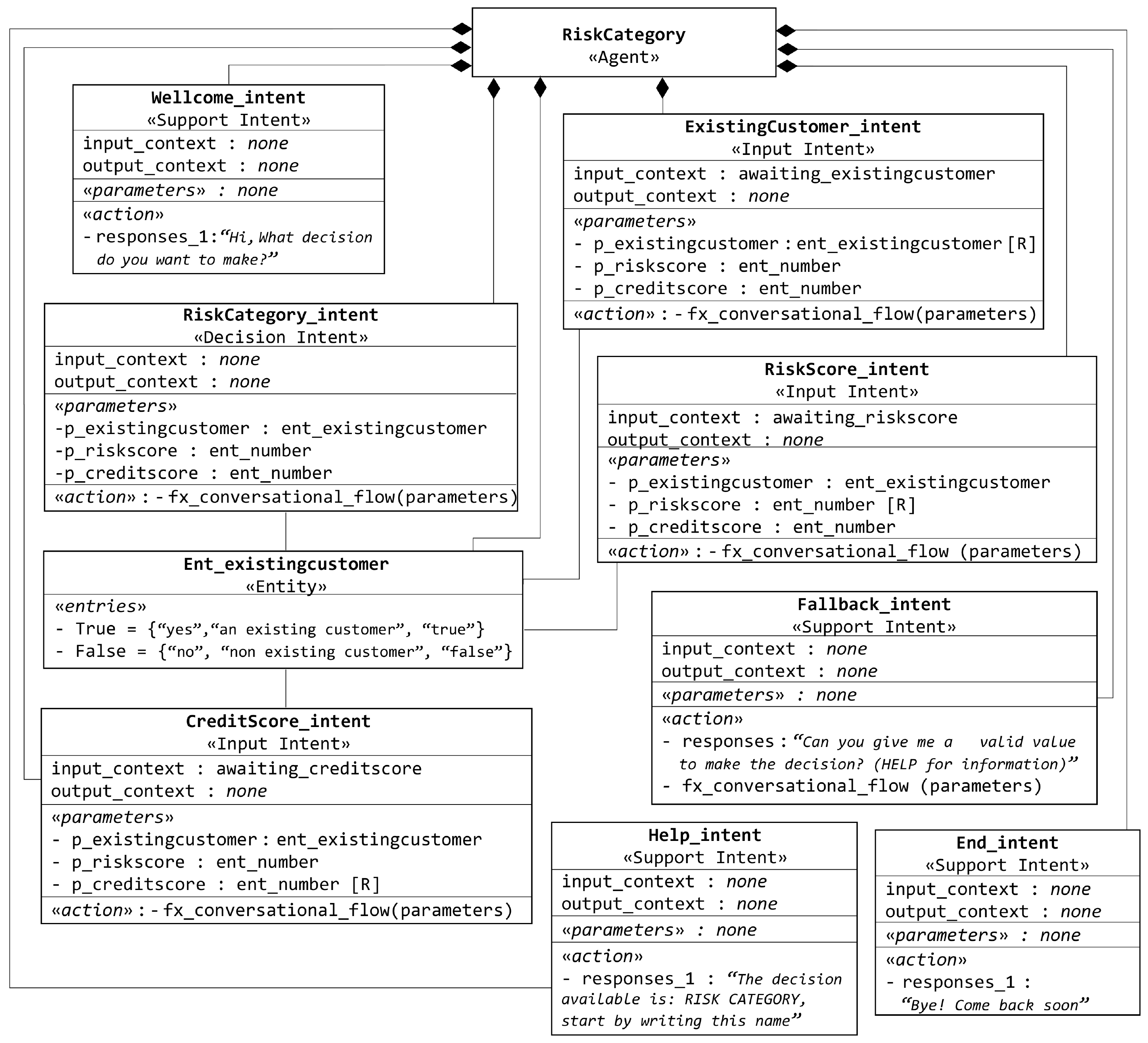}
	\caption{Elements of the agent resulting from the methodology.}
	\label{fig:elementsdiagram}       
\end{figure}

To illustrate the result of the methodology described in the mapping of the previous steps, Figure \ref{fig:elementsdiagram} shows the elements of the resulting chatbot agent for the model described in Figure \ref{fig:dmn_table_1} or \ref{fig:dmn_table_2}. It includes the entity for the existent customer input and the decision, input and support intents.

\subsection{Including Multiple Decisions in the Same Chatbot}
\label{subsec:methodology_multiple_decisions}

Our methodology is designed to allow the use of only one chatbot to support multiple decisions. To this end, it is necessary to consider two different situations depending on the relationship between the decisions. 

If decisions are independent of each other, it is enough to repeat the same methodology for each decision and add the resulting intents, entities and actions to the same chatbot. The only thing that needs to be considered is to add some prefix when generating all the names in order to avoid a name clash between them. When the conversation with the chatbot starts, the chatbot will try to match the user input with all of the decision intents, which are the only intents without input context. Once a decision intent is matched, the context of the chosen decision will be set. This will prevent other decision intents or input intents of different decisions from being matched with user inputs. 

If there is a hierarchical relationship between the decisions, i.e., when one of the inputs of a decition table is the result of another decision table, there is no need to apply the methodology to each decision separately. Instead, it is enough to apply it to the decision in the top of the hierarchy. Only two things need to be considered in this case. First, the set of inputs used to build the chatbot must be the union of the inputs of all the decision tables that belong to the hierarchy. And second, function \texttt{is\_necessary} needs to be extended to consider that inputs could be found in different decision tables. 
\section{Implementation, Evaluation and Limitations}
\label{sec:evaluation}

The proposed methodology is not restricted to a single tool for generating chatbots. In this article we use $Dialogflow$, which is a tool for generating chatbots and communication interfaces that allows text and voice conversations, based on artificial intelligence and that offers the possibility of connecting with other existing communication applications. 
A set of chatbots were built based on different DMN decision models. Examples include simple decisions, involving decisions with known values, as well as decisions that use wildcards (See examples in Figures~\ref{fig:dmn_table_1}, \ref{fig:dmn_table_2} and \ref{fig:dmn_table_3}); and hierarchical decisions where the value of an input is the result of another decision.
To facilitate the interactions between the chatbot and the user, integration with Telegram and a web demo provided by Dialogflow were performed. Chatbot examples are available online\footnote{\url{https://github.com/Adartse/DMNChatbots}}. For each example we provide the chatbot source code, its link for the web version, sample conversations in Telegram and the DMN model that each conversation solves.

To evaluate the proposed methodology, chatbots were built following the methodology. Then, the opinions of 17 people from academia and industry were collected through online questionnaires. We were interested in knowing the perception of the usefulness of the result and the opinion about the use experience. In this way we evaluate, for example, if they consider the support attempts useful, or the phrases used to issue the response actions. The following is a list of the questions included in the evaluation questionnaire.

\begin{description}
\item[$Q1:$] \textit{How was your interaction with the chatbot?}
\item[$Q2$:] \textit{How useful did you find the assistance provided by the chatbot in case you asked for it? (only answer if you used it)}
\item[$Q3$:] \textit{Considering the decision table below, did the decision-support chatbot ask you the right questions to come to a decision?} 
\item[$Q4$:] \textit{Do you see potential for this kind of applications in organizations?} 
\item[$Q5$:] \textit{What did you like/dislike about the chatbot?}
\item[$Q6$:] \textit{Do you have any suggestions to improve the Decision-Support Chatbot?}
\end{description}

The full details of the results are available online$^6$. With respect to the first four questions, which have a linear scale (from 1 to 4), we get that: 

\begin{itemize}

\item For $Q1$, 47\% considered that they had a fluent or very fluent conversation with the chatbot. Only 13\% considered the conversation not fluid at all.
\item For $Q2$, 93\% used some support command during the conversation, of which 61\% consider that they received useful information.
\item For $Q3$, Only 10\% of the respondents considered that the chatbot did not ask the right questions to make the decision, another 10\% considered that it did so poorly, and the remaining 80\% considered that it did so correctly or absolutely correctly.
\item And for $Q4$, 80\% considered that this type of application has potential or large potential in organizational contexts.

\end{itemize}

The answers to $Q5$ and $Q6$ are closely related. Although we received positive feedback highlighting the usefulness of this type of tools, especially for more complex decision making scenarios, the main comments on what people disliked and should be improved 
were focused on (i) providing more information on the context of the decision to be made and (ii) expanding and customizing the help options during the conversation. 
To solve (i) during transformation process, additional information on the decisions should be requested in order to build the context information intents as new Support Intents. And with regard to (ii), a very similar solution is proposed: collect information on the type and range of values expected for each input and build in a set of additional Support Intents. 
Both comments will serve as a basis to further expand the methodology proposed here to encourage increasingly friendly conversations and will be the basis for developing tools to automate the transformation of DMN models into chatbots. 

\section{Conclusions and Future Work}
\label{sec:conclusions_future_work}

In this article we introduced a novel methodology that allows the creation of chatbots, in a semi-automatic way, through the systematic transformation of DMN decision tables. The chatbots built following these steps are able to ask for the required inputs to evaluate the decision rules defined to obtain a concrete decision output. This is done reducing the number of interactions with the user, for example when wildcards are included in the decision definition or collecting information about several inputs at the same time. In addition, the methodology provides the opportunity of reusing domain-independent parts of chatbots, like those provided by support intents, in all decision chatbots.

The generated chatbots are prototypes and as it is derived from the evaluation, although in most cases satisfactory conversations are obtained that show an adequate response to the provided inputs, there are still functionalities that can be improved or extended, such as providing more support to the user during the conversation by allowing questions about the already provided inputs, rectifying previously given values, raising what-if scenarios, or adding patterns to the automatic sentence generation to improve the users' ability to detect sentences. However, it should be remembered that the aim of this article is not to provide a production-ready chatbot, but to prove that it is possible to generate it semi-automatically from DMN models. 

In addition to enriching interactions with the user, we foresee the need to continue working to achieve the automatic transformation of DMN models defined in a standard way, for example using XML files.


\bibliographystyle{ieeetr}
\bibliography{bibfile}

\end{document}